\documentclass[AMA,STIX2COL,Linenumbersoff]{MRM}
\articletype{Research Article}%
\usepackage{amsmath,amssymb,amsfonts}
\usepackage{graphicx}
\usepackage{textcomp}
\usepackage{xspace}
\usepackage{bm}
\long\def\comment#1{}

\newcommand{\citestd}[1] {Ref.~\citen{#1}\xspace}
\newcommand{\fref}[1] {Fig.~\ref{#1}\xspace} % the non-breaking space "~" is very important!  use macros!
\newcommand{\tref}[1] {Table~\ref{#1}\xspace}
\newcommand{\kspace}{k-space\xspace}

\newcommand{\xmath}[1] {\ensuremath{#1}\xspace}
\newcommand{\blmath}[1] {\xmath{\bm{#1}}}

\newcommand{\resp}[1]{}

\newcommand{\omdd}[1] {\xmath{\bm{\omega}^{[#1]}}}
\newcommand{\omd} {\omdd{d}}

\newcommand{\rd} {\xmath{\bm{r}^{[d]}}}

\newcommand{\etat} {\xmath{\eta_{\theta}}}
\newcommand{\etao} {\xmath{\eta_{\om}}}

\newcommand{\cc} {\xmath{\bm{c}}}

\newcommand{\argmin} {\operatornamewithlimits{arg\,min}} % argmin

\newcommand{\diag}[1] {\mathrm{diag}\!\left\{#1\right\}}

\newcommand{\reals} {\xmath{\mathbb{R}}}

\newcommand{\A} {\blmath{A}}
\newcommand{\B} {\blmath{B}}
\newcommand{\D} {\blmath{D}}

\newcommand{\T} {\blmath{T}}

\newcommand{\x} {\blmath{x}}
\newcommand{\xh} {\xmath{\hat{\x}}}

\newcommand{\gi} {\blmath{g}_i}
\newcommand{\pid} {\blmath{p}_{id}}
\newcommand{\si} {\blmath{s}_i}
\newcommand{\sid} {\blmath{s}_{id}}
\newcommand{\y} {\blmath{y}}

\newcommand{\vveps} {\blmath{\varepsilon}}
\newcommand{\om} {\blmath{\omega}}
\newcommand{\thta} {\blmath{\theta}}

\newcommand{\Loss}{\mathcal{L}}
\newcommand{\Lrecon}{\Loss_{\mathrm{recon}}}
\newcommand{\Lg}{\Loss_{\mathrm{g}}}
\newcommand{\Ls}{\Loss_{\mathrm{s}}}
\newcommand{\Lpns}{\Loss_{\mathrm{pns}}}

\newcommand{\gmax} {\xmath{g_{\max}}}
\newcommand{\smax} {\xmath{s_{\max}}}
\newcommand{\smin} {\xmath{s_{\min}}}
\newcommand{\pmax} {\xmath{p_{\max}}}
\newcommand{\dt} {\xmath{\Delta t}}

\newcommand{\Ns} {\xmath{N_{\mathrm{s}}}} % # samples
\newcommand{\Nfe} {\xmath{N_{\mathrm{fe}}}} % # frequency encoding
\newcommand{\Nd} {\xmath{N_{\mathrm{d}}}} % # dimensions
\long\def\red#1{\bgroup\color{red} #1 \egroup}
\received{}
\revised{}
\accepted{}
\begin{document}

\title{Stochastic Optimization of 3D Non-Cartesian Sampling
Trajectory (SNOPY) \protect\thanks{This manuscript is submitted to \textit{Magnetic Resonance in Medicine} on 09/15/22.
}}

\author[1]{Guanhua Wang}{\orcid{0000-0002-1622-5664}}

\author[1]{Jon-Fredrik Nielsen }{\orcid{0000-0002-2058-3579}}

\author[1,2]{Jeffrey A. Fessler}{\orcid{0000-0001-9998-3315}}

\author[1]{Douglas C. Noll}{\orcid{0000-0002-0983-3805}}

\authormark{Wang \textsc{et al}}

\address[1]{\orgdiv{Biomedical Engineering}, \orgname{University of Michigan}, \orgaddress{\state{Michigan}, \country{United States}}}

\address[2]{\orgdiv{EECS}, \orgname{University of Michigan}, \orgaddress{\state{Michigan}, \country{United States}}}

\corres{Guanhua Wang. \email{guanhuaw@umich.edu}}

\finfo{This work was partially supported by
\fundingAgency{National Science Foundation} grant \fundingNumber{IIS 1838179}
and \fundingAgency{National Institutes of Health} grant \fundingNumber{R01 EB023618, U01 EB026977}.}

\abstract{
\section{Purpose} 
Optimizing 3D k-space sampling trajectories
for efficient MRI
is important yet challenging.
This work
proposes a generalized framework
for optimizing 3D non-Cartesian sampling patterns via 
data-driven optimization.

\section{Methods} 
We built a differentiable MRI system model to enable
gradient-based methods for
sampling trajectory optimization.
By combining training losses,
the algorithm can simultaneously
optimize multiple properties of sampling patterns,
including image quality, 
hardware constraints (maximum slew rate and gradient strength),
reduced peripheral nerve stimulation (PNS),
and parameter-weighted contrast.
The proposed method can either
optimize the gradient waveform
(spline-based freeform optimization)
or optimize properties of given sampling trajectories
(such as the rotation angle of radial trajectories).
Notably, the method optimizes sampling trajectories
synergistically with either model-based 
or learning-based reconstruction methods.
We proposed several strategies to alleviate
the severe non-convexity and huge computation demand
posed by the high-dimensional optimization.
The corresponding code is organized
as an open-source, easy-to-use toolbox.

\section{Results}
We applied the optimized trajectory to multiple applications
including structural and functional imaging.
In the simulation studies,
the reconstruction PSNR of a 3D kooshball trajectory
was increased by 4 dB with SNOPY optimization.
In the prospective studies,
by optimizing the rotation angles of a stack-of-stars (SOS) trajectory,
SNOPY improved the PSNR by 1.4 dB compared to the best empirical method.
Optimizing the gradient waveform of a rotational EPI trajectory
improved subjects’ rating of the PNS effect from ‘strong’ to ‘mild.’

\section{Conclusion} SNOPY provides an efficient data-driven 
and optimization-based method
to tailor non-Cartesian sampling trajectories.}

\keywords{Magnetic resonance imaging, non-Cartesian sampling,
data-driven optimization, image acquisition, deep learning}

% \jnlcitation{\cname{%
% \author{Williams K.}, 
% \author{B. Hoskins}, 
% \author{R. Lee}, 
% \author{G. Masato}, and 
% \author{T. Woollings}} (\cyear{2016}), 
% \ctitle{A regime analysis of Atlantic winter jet variability applied to evaluate HadGEM3-GC2}, \cjournal{Q.J.R. Meteorol. Soc.}, \cvol{2017;00:1--6}.}

\maketitle

\footnotetext{
\textbf{Abbreviations:}~\hbox{MRI,~magnetic~resonance~imaging;
PSNR,~peak~signal-to-noise} ratio}

\section{Introduction}\label{sec:intro}

Most magnetic resonance imaging systems
sample data in the frequency domain (\kspace)
following prescribed sampling trajectories.
Efficient sampling strategies can
accelerate acquisition and improve image quality.
Many well-designed sampling strategies and their variants,
such as spiral, radial, CAIPIRINHA, and PROPELLER
\cite{ahn1986spiral, lauterbur1973image, breuer:2006:caipi, pipe:1999:PROPELLER},
have enabled MRI's application to many areas \cite{glover:2001:SpiralinOutBOLD,johansson:2018:RigidbodyMotionCorrection, yu:2013:ClinicalApplicationControlled, forbes:2001:PROPELLERMRIClinical}.
Sampling patterns in \kspace
either are located on the Cartesian raster
or arbitrary locations (non-Cartesian sampling).
This paper focuses on optimizing 3D 
non-Cartesian trajectories
and introduces a generalized 
gradient-based optimization method
for automatic trajectory design/tailoring.

The design of sampling patterns usually considers
certain properties of \kspace signals.
For instance, the variable density (VD) spiral trajectory \cite{lee:2003:Fast3DImaging}
samples more densely in the central \kspace
where more energy is located.
For higher spatial frequency regions,
the VD spiral trajectory uses larger gradient strengths
and slew rates
to cover as much of \kspace as quickly as possible.
Compared to 2D sampling,
designing 3D sampling by hand
is more challenging for several reasons.
The number of parameters increases in 3D,
and thus the parameter selection is more difficult 
due to the larger search space.
Additionally, analytical designs usually are
based on the Shannon-Nyquist relationship \cite{gurney:2006:DesignAnalysisPractical,johnson:2017:HybridRadialconesTrajectory,zhou:2017:GoldenratioRotatedStackofstars}
that might not fully consider
properties of sensitivity maps
and advanced reconstruction methods.
For 3D sampling pattern with a high undersampling (acceleration) ratio,
there are limited analytical tools
for designing sampling patterns
having anisotropic FOV and resolution.
The peripheral nerve stimulation (PNS) effect
is also more severe in the 3D imaging \cite{ham:1997:PeripheralNerveStimulation},
further complicating analytical trajectory designs.
For these reasons, it is important to automate the design of 3D sampling trajectories.

Many 3D sampling approaches exist.
`stack-of-2D' is a commonly used 3D sampling strategy,
by stacking 2D sampling patterns
in the slice direction \cite{johansson:2018:RigidbodyMotionCorrection,ham:1997:PeripheralNerveStimulation}.
This approach is easier to implement 
and enables slice-by-slice 2D reconstruction.
Another design applies Cartesian sampling in the frequency-encoding direction
and non-Cartesian sampling in the phase-encoding direction \cite{aggarwal:2020:JointOptimizationSampling,bilgic:2015:WaveCAIPIHighlyAccelerated}.
However, these approaches do not
fully exploit the potential of modern gradient systems
and may not perform as well as true 3D sampling trajectories
\cite{chaithya:2022:OptimizingFull3D}. 

Recently, 3D SPARKLING \cite{chaithya:2022:OptimizingFull3D}
proposes to optimize 3D sampling trajectories
based on 
the goal of
conforming to a given density
while distributing samples as uniformly as possible \cite{sparklingmrm}.
That method demonstrated improved image quality
compared to the `stack-of-2D SPARKLING' approach.
In both 2D and 3D,
the SPARKLING approach
uses a pre-specified sampling density in \kspace
that is typically 
an isotropic radial function.
This density function
cannot readily capture distinct energy distributions
of different imaging protocols.
Additionally, the method does not control PNS effects explicitly.
SPARKLING optimizes the location of every sampling point,
or the gradient waveform
(freeform optimization),
and is not applicable to the optimization of
existing parameters of
existing sampling patterns,
which limits its practicability beyond
T2$^*$-weighted imaging.

In addition to analytical methods,
learning-based methods are also investigated
in MRI sampling trajectory design.
Since different anatomies have distinct energy
distributions in the frequency domain,
an algorithm may learn to optimize sampling trajectories 
from training datasets.
Several studies show that different anatomies
and different reconstruction algorithms
produce very different optimized sampling patterns,
and such optimized sampling trajectories can improve image quality \cite{huijben:2020:LearningSamplingModelBased, wang:22:bjork-tmi,bahadir:2020:DeepLearningBasedOptimizationUnderSampling, pilot, sanchez:2020:ScalableLearningBasedSampling,jin:2019:SelfSupervisedDeepActive,
rl:david,sherry:20:lts,gozcu2018learning}.
Several recent studies also applied learning-based approaches
to 3D non-Cartesian trajectory design.
J-MoDL \cite{aggarwal:2020:JointOptimizationSampling}
proposes to learn sampling patterns and model-based deep learning reconstruction algorithms jointly.
J-MoDL optimizes the sampling locations in the phase-encoding direction,
to avoid the computation cost 
of non-uniform Fourier transform.
PILOT/3D-FLAT \cite{alush-aben:2020:3DFLATFeasible}
jointly optimizes freeform 3D non-Cartesian trajectories
and a reconstruction neural network
with gradient-based methods. 
These studies use the standard %brute-force
auto-differentiation approach to calculate the gradient,
which can be inaccurate and 
lead to sub-optimal optimization results \cite{wang:22:bjork-tmi,wang:21:eao}.

This work extends our previous methods
\cite{wang:22:bjork-tmi, wang:21:eao}
and introduces a generalized
\textbf{S}tochastic optimization framework
for 3D \textbf{NO}n-Cartesian sam\textbf{P}ling trajector\textbf{Y} (\textbf{SNOPY}).
The proposed method can automatically tailor given trajectories
and learn \kspace features from training datasets.
We formulated several optimization objectives,
including image quality, hardware constraints,
PNS effect suppression, and image contrast.
Users can simultaneously optimize one or multiple characteristics
of a given sampling trajectory.
Similar to previous learning-based methods \cite{aggarwal:2020:JointOptimizationSampling,
wang:22:bjork-tmi,bahadir:2020:DeepLearningBasedOptimizationUnderSampling,pilot},
the sampling trajectory can be jointly optimized
with trainable reconstruction algorithms
to improve image quality.
The joint optimization approach
can exploit the synergy between
acquisition and reconstruction,
and learn optimized trajectories 
for different anatomies.
The algorithm can optimize various properties of a sampling trajectory,
such as the readout waveform, or the rotation angles of readout shots,
making it more practical and applicable.
We also introduced several techniques to
improve efficiency,
enabling large-scale 3D trajectory optimization.
We tested the proposed methods
with multiple imaging applications,
including structural and functional imaging.
These applications benefited from the
SNOPY-optimized sampling trajectories
in both simulation and prospective studies.

\section{Theory}
\label{sec:method}
\begin{figure*}
\centerline{\includegraphics[width=0.8\textwidth]{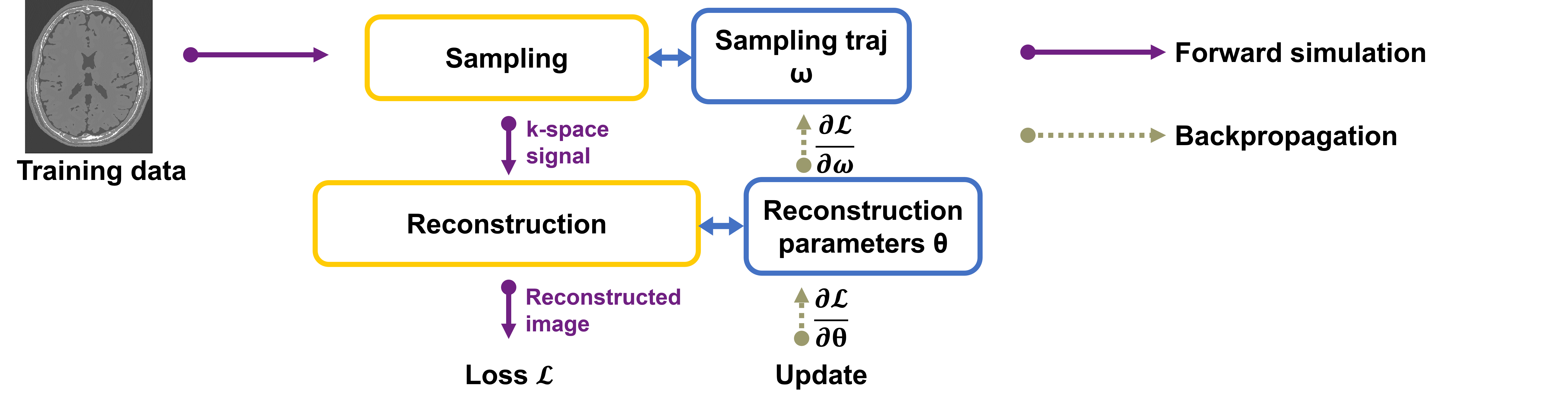}}
\caption{~Diagram of SNOPY.
The sampling trajectory (and possibly reconstruction parameters)
are updated using gradient methods.
The training/optimization uses the differentiable programming approach
to get the gradient required in the update.
\label{fig:workflow}}
\end{figure*}

This section describes the proposed gradient-based methods for 
trajectory optimization.
We use the concept of differentiable programming
to compute the Jacobian/gradient
w.r.t. sampling trajectories required
in the gradient-based methods.
The sampling trajectory and reconstruction parameters
are differentiable parameters,
whose gradients can be calculated
by auto-differentiation or chain rule.
To learn/optimize these parameters,
one may apply (stochastic) gradient descent-like algorithms.
\fref{fig:workflow} illustrates the basic idea.
Here the sampling trajectory can also be jointly optimized with
the parameters of a learnable reconstruction algorithm,
so that the learned sampling trajectory and reconstruction method
are in synergy and can produce high-quality images.
The SNOPY algorithm combines
several optimization objectives,
to ensure that the optimized sampling trajectories 
have desired properties.
\ref{subsec:obj} delineates these objective functions.
\ref{subsec:param} shows that the proposed method
is applicable to multiple scenarios with
different parameterization strategies.
For non-Cartesian sampling,
the system model usually involves non-uniform fast Fourier transforms (NUFFT).
\ref{subsec:jacob} briefly describes an
efficient and accurate way to calculate the gradient involving NUFFTs.
\ref{subsec:efficient}
suggests several engineering approaches to 
make this large-scale optimization problem practical and efficient.

\subsection{Optimization objectives}
\label{subsec:obj}

This section describes the
optimization objectives. 
Since we propose to use a stochastic gradient descent-type
optimization algorithm,
the objective function, or loss function,
is by default defined on a mini-batch of data.
The final loss function can be a linear combination
of the following loss terms
to ensure the optimized trajectory
has multiple desired properties.

\subsubsection{Image quality}

For many MRI applications,
efficient acquisition and reconstruction aim to produce high-quality images.
Consequently, the learning objective should encourage
images reconstructed
from sampled \kspace signals to be close to the reference image.
We formulate this similarity objective
as the following image quality training loss:
\begin{equation}
    \label{eqn:image}
\Lrecon = 
\|f_{\thta, \om}(\A(\om)\x+\vveps) - \x\|.
\end{equation}
Here,
$\om(\cc) \in \reals^{\Nfe \times \Ns \times \Nd}$
denotes the trajectory to be optimized,
with \Ns shots, \Nfe sampling points in each shot, and \Nd image dimensions.
\cc denotes the parameterization coefficients introduced in \ref{subsec:param}.
For 3D MRI, $\Nd=3$.
\x is a mini-batch of data from the training set $\mathcal{X}$.
\vveps is simulated complex Gaussian noise.
$\A(\om)$ is the forward system matrix for sampling trajectory \om \cite{fessler:03:nff}.
\A can also incorporate multi-coil sensitivity information \cite{pruessmann:2001:AdvancesSensitivityEncoding}
and field inhomogeneity \cite{fessler:05:tbi}.
\thta denotes the reconstruction algorithm's parameters.
It can be kernel weights in a convolutional neural network (CNN),
or the regularizer coefficient in a model-based reconstruction method.
The term $\|\cdot\|$ can be $\ell_1$ norm,
$\ell_2$ norm,
or a combination of both.
There are also other ways to measure the distance between $\x$ and
$f_{\thta, \om}(\A(\om)\x+\vveps)$,
such as the structural similarity index (SSIM \cite{wang:2004:ssim}).
For simplicity,
this work used a linear combination of $\ell_1$ norm and square-of-$\ell_2$ norm.

\subsubsection{Hardware limits}

The gradient system of MR scanners
has physical constraints,
namely maximum gradient strength and slew rate.
Ideally, we would like to enforce
a set of constraints of the form
\[
\| \gi[j,:] \|_2 \leq g_{\mathrm{max}}
,\quad
\gi = \D_1 \om[:,i,:] / (\gamma \dt) \in \reals^{(\Nfe-1) \times \Nd}
,\]
for every shot $i = 1,\ldots,\Ns$
and time sample $j = 1,\ldots,\Nfe$,
where $\gi$ denotes the gradient strength
of the $i$ shot
and
\gmax
denotes the desired gradient upper bound.
We use a Euclidean norm along the spatial axis
so that any 3D rotation of the sampling trajectory 
still obeys the constraint.
A similar constraint is enforced
on the Euclidean norm
of the slew rate
\(
\si = \D_2 \om[:,i,:] / (\gamma \dt^2)
,\)
where
$\D_1$ and $\D_2$
denote first-order and second-order
finite difference operators applied along the readout dimension,
and
\dt is the raster time interval
and $\gamma$ is the gyromagnetic ratio.

To make the optimization more practical,
we follow previous studies
\cite{wang:22:bjork-tmi,pilot},
and formulate the hardware constraint as a soft penalty term:
\begin{equation}
    \label{eqn:grad}
    \Lg=
    \sum_{i=1}^{\Ns} \sum_{j=1}^{\Nfe-1} 
    \phi_{\gmax}(\| \gi[j,:] \|_2)
\end{equation}
\begin{equation}
    \label{eqn:slew}
    \Ls=
    \sum_{i=1}^{\Ns} \sum_{j=1}^{\Nfe-2} 
    \phi_{\smax}(\| \si[j,:] \|_2)
.\end{equation}
Here
$\phi$ is a penalty function,
and we use a soft-thresholding function
$\phi_\lambda(x)
= \max(|x|-\lambda, 0).$
Since $\phi$ here is a soft penalty
and the optimization results may exceed \smax and \gmax,
\smax and \gmax can be slightly lower than
the scanner's physical limits
to make the optimization results feasible on the scanner.

\subsubsection{Suppression of PNS effect}

3D imaging often
leads to stronger PNS
effects than 2D imaging
because of the additional gradient
axis.
To quantify possible PNS effects
of candidate gradient waveforms,
we used the convolution model in
\citestd{schulte:2015:PNS}:
\begin{equation}
    R_{id}(t)=\frac{1}{\smin} \int_0^t \frac{\sid(\theta) c}{(c+t-\theta)^2} d \theta,
\end{equation}
where $R_{id}$ denotes the PNS effect of the gradient waveform from the $i$th shot and the $d$th dimension.
$\sid$ is the slew rate of the $i$th shot in the $d$th dimension.
\xmath{c} (chronaxie) and \smin (minimum stimulation slew rate)
are scanner parameters.

Likewise, we discretize the convolution model
and formulate a soft penalty term 
as the following loss function:
\begin{equation} 
\pid[j] = \sum_{k=1}^{j}\frac{\sid[j] c  \dt}{\smin(c + j \dt - k \dt)^2},
\nonumber
\end{equation}
\begin{equation}
    \label{eqn:pns}
    \Lpns=
    \sum_{i=1}^{\Ns} \sum_{j=1}^{\Nfe} \phi_{\pmax}((\sum_{d=1}^{\Nd}\pid[j]^2)^{\frac{1}{2}}).
\end{equation}

Again, $\phi$ denotes the soft-thresholding function,
with PNS threshold \pmax (usually $\leq 80\%$\cite{schulte:2015:PNS}). 
This model combines the 3 spatial axes via the sum-of-squares manner, 
and does not consider the anisotropic response of PNS \cite{davids2019prediction}.
The implementation may use an FFT (with zero padding)
to implement this convolution efficiently.

\subsubsection{Image contrast}

In many applications,
the optimized sampling trajectory should 
maintain certain parameter-weighted contrasts.
For example,
ideally the (gradient) echo time (TE)
should be identical for each shot.
Again
we replace this hard constraint 
with an echo time penalty.
Other parameters, like repetition time (TR)
and inversion time (TI),
can be predetermined in the RF pulse design.
Specifically, the corresponding loss function
encourages the sampling trajectory to cross
the \kspace center at certain time points:
\begin{equation}
    \label{eqn:contrast}
    \Loss_c =
    \sum_{\{i,j,d\} \in C}
    \phi_{0}(|\om[i,j,d]|), 
\end{equation}

where $C$ is a collection of gradient 
time points that are constrained 
to cross k-space zero point.
$\phi$ is still a
soft-thresholding function,
with threshold 0.

\subsection{Reconstruction}
In \eqref{eqn:image},
the reconstruction algorithm $f_{\thta, \om}(\cdot)$
can be various algorithms.
Consider a typical cost function
for regularized MR image reconstruction
\begin{equation}
\label{recon}
\xh = \argmin_{\x} \|\A(\om)\x-\y\|_{2}^2 + \mathcal{R}(\x).
\end{equation}
$\mathcal{R}(\x)$ here can be a Tikhonov regularization $\lambda \|\x\|_2^2$
(CG-SENSE \cite{maier:2021:CGSENSERevisitedResults}),
a sparsity penalty $\lambda \|\T\x\|_1$ 
(compressed sensing \cite{lustig:2008:CompressedSensingMRI}, $\T$ is a finite-difference operator),
a roughness penalty $\lambda \|\T\x\|_2^2$ 
(penalized least squares, PLS),
or a neural network (model-based deep learning, MoDL \cite{modl}).
The Results section shows that different reconstruction algorithms
lead to distinct optimized sampling trajectories.

To get a reconstruction estimation $\xh$, 
one may use corresponding iterative reconstruction algorithms.
Specifically, the algorithm should be step-wise
differentiable (or sub-differentiable)
to enable differentiable programming.
The backpropagation
uses the chain rule to traverse 
every step of the iterative algorithm
to calculate the gradient w.r.t. variables
such as \om.

\subsection{Parameterization}
\label{subsec:param}
As is shown in \citestd{wang:22:bjork-tmi},
directly optimizing
every \kspace sampling point
(or equivalently every gradient waveform time point)
may lead to sub-optimal results.
Additionally, in many applications one wants to
optimize certain
properties of existing sampling patterns,
such as the rotation angles of a multi-shot spiral trajectory,
so that the optimized trajectory can 
be easily integrated into existing workflows.
For these cases,
we propose two parameterization strategies.

The first approach,
spline-based freeform optimization, 
is to represent the sampling pattern
using a linear basis,
i.e.,
$\om = \B \cc$,
where \B is a matrix of samples
of a basis
such as quadratic B-spline kernels
and \cc denotes
the coefficients to be optimized
\cite{wang:22:bjork-tmi,pilot}.
This approach fully exploits the generality
of a gradient system.
Using a linear parameterization like B-splines
reduces the degrees of freedom
and facilitates applying hardware constraints \cite{wang:22:bjork-tmi,hao:2016:JointDesignExcitationa}.
Additionally, it enables multi-scale optimization 
for avoiding sub-optimal local minima
and further improving optimization results \cite{sparklingmrm,wang:22:bjork-tmi,pilot}.
However, the freeformly optimized trajectory could have implementation challenges. For example, some MRI systems can not restore hundreds of different gradient waveforms. 

The second approach is to optimize attributes (\cc) of existing trajectories
such as rotation angles,
where $\om(\cc)$ is a nonlinear function of the parameters.
The trajectory parameterization
should be differentiable in \cc
to enable differentiable programming.
This approach is easier to implement on scanners,
and can work with existing workflows, such as reconstruction and eddy-current correction,
with minimal modification.

\subsection{Efficient and accurate Jacobian calculation}
\label{subsec:jacob}
In optimization,
the sampling trajectory is
embedded in the forward system matrix
within the similarity loss \eqref{eqn:image}.
The system matrix for non-Cartesian sampling
usually includes a NUFFT operation
\cite{fessler:03:nff}.
Updating the sampling trajectory
in each optimization step
requires the Jacobian, or the gradient w.r.t.
the sampling trajectory.
The NUFFT operator typically involves interpolation 
in the frequency domain,
which is non-differentiable in typical implementations
due to rounding operations.
Several previous works used auto-differentiation
(with sub-gradients)
to calculate an approximate
numerical gradient \cite{pilot,alush-aben:2020:3DFLATFeasible},
but that approach is inaccurate and slow \cite{wang:21:eao}.
We derived an efficient and accurate
Jacobian approximation method \cite{wang:21:eao}.
For example,
the efficient Jacobian of a forward system model \A is:
\begin{equation}
\frac{\partial \A\x}{\partial \omd} = -\imath \, \diag{\A (\x\odot\rd)}
\label{e,Ax}
,\end{equation}
where $d \in \{1,2,3\}$ denotes a spatial dimension,
and $\rd$ denotes the Euclidean spatial grid.
Calculating this Jacobian
simply uses another NUFFT, 
which is more efficient than the auto-differentiation approach.
See 
\citestd{wang:21:eao} for more cases,
such as $\frac{\partial \A'\A\x}{\partial \omd}$ and the detailed derivation.

\subsection{Efficient optimization}
\label{subsec:efficient}
\subsubsection{Optimizer}

\begin{table*}[t]%
\caption{~The memory/time use reduction brought by proposed techniques.
Here we used a 2D 400$\times$400 test case,
and CG-SENSE reconstruction (20 iterations). `+' means adding the technique to previous columns.
}
\label{tab:memory}
\begin{tabular*}{\textwidth}{@{\extracolsep\fill}lcccc@{\extracolsep\fill}}
\toprule
\textbf{Plain} & \textbf{+Efficient Jacobian}  & \textbf{+In-place ops}  & {\textbf{+Toeplitz embedding}}  & \textbf{+Low-res NUFFT}   \\
\midrule
5.7GB / 10.4s  &  272MB / 1.9s  & 253MB / 1.6s  & 268MB / 0.4s  & 136MB / 0.2s  \\
\bottomrule
\end{tabular*}
\begin{tablenotes}%%[341pt]
\end{tablenotes}
\end{table*}

Generally, to optimize the sampling trajectory $\om$
and other parameters (such as reconstruction parameters $\thta$) via
stochastic gradient descent-like methods,
each update needs to take a gradient step
(in the simplest form)
\begin{align*}
\thta^K &= \thta^{K-1} - \etat  \frac{\partial \Loss}{\partial\thta}(\om^{K-1}, \thta^{K-1})
\\
\om^K &= \om^{K-1} - \etao  \frac{\partial \Loss}{\partial\om}(\om^{K-1}, \thta^{K-1}),
\end{align*}
where $\Loss$ is the loss function described
in Section \ref{subsec:obj} 
and
%$\eta$ is the step size.
where \etat and \etao denote step-size parameters.

The optimization is highly non-convex
and may suffer from sub-optimal local minima.
We used stochastic gradient Langevin dynamics (SGLD)
\cite{welling:2011:sgld} as the optimizer to improve results and
accelerate training.
Each update of SGLD injects Gaussian noise into the gradient
to introduce randomness
\begin{align}
\thta^K &= \thta^{K-1} - \etat
\frac{\partial \Loss}{\partial\thta^{K-1}} + \sqrt{2 \eta_{\thta}} \mathcal{N}(0,\,1)
\nonumber \\
\om^K &= \om^{K-1} - \etao
\frac{\partial \Loss}{\partial\om^{K-1}} + \sqrt{2\etao} \mathcal{N}(0,\,1)
\label{e:sgld}
.\end{align}

\begin{figure}[htb]
\centerline{\includegraphics[width=\columnwidth]{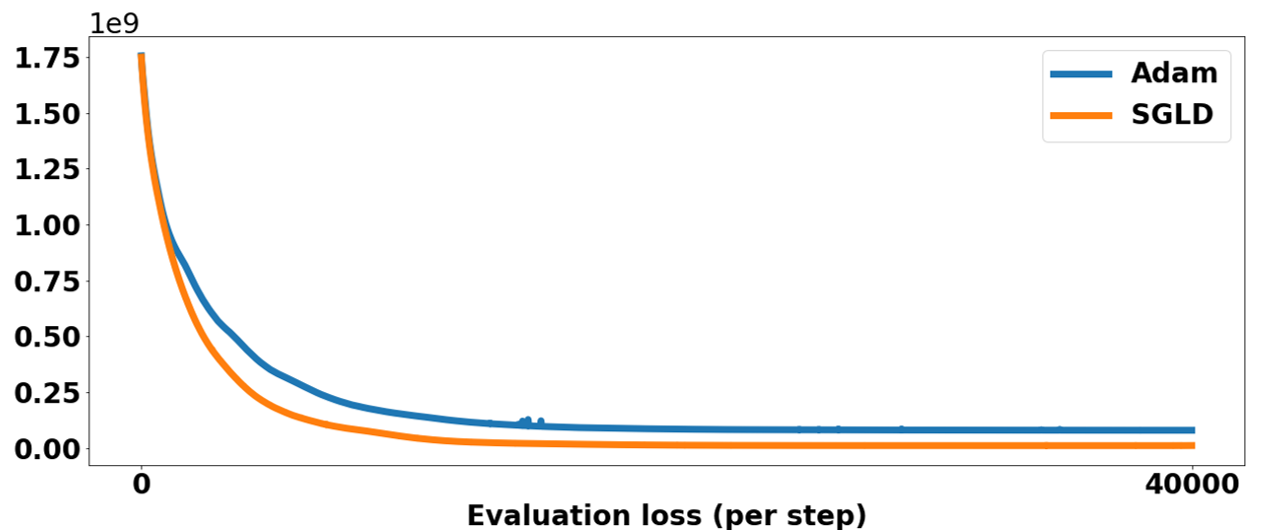}}
\caption{~The evaluation loss curve led by SGLD and Adam.}
\label{fig:loss}
\end{figure}

Across most experiments, 
we observed that SGLD led to improved results 
and better convergence speeds
compared with SGD or Adam \cite{kingma:2017:AdamMethodStochastic}.
\fref{fig:loss} shows a loss curve of SGLD
and Adam of experiment \ref{exp:pns}.

\begin{figure*}[htb]
\centerline{\includegraphics[width=0.9\textwidth]{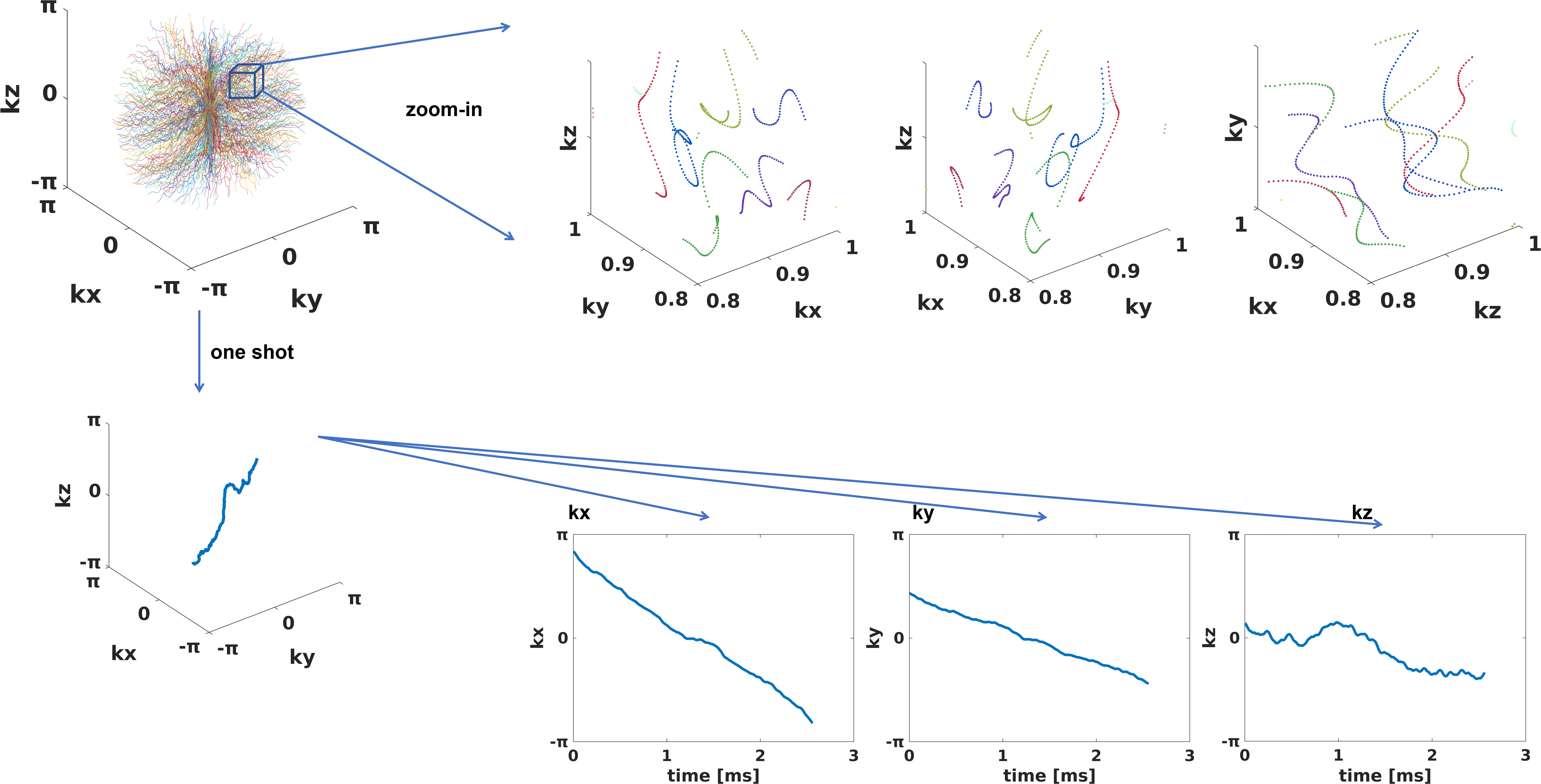}}
\caption{~The optimized sampling trajectory of experiment \ref{exp:freeform}.
The training involves SKM-TEA dataset and MoDL \cite{modl} reconstruction.
The upper row shows a zoomed-in region from different viewing perspectives.
The lower row shows one shot from different perspectives.
\label{fig:bjork}}
\end{figure*}

\begin{figure}[htb]
\centerline{\includegraphics[width=0.9\columnwidth]{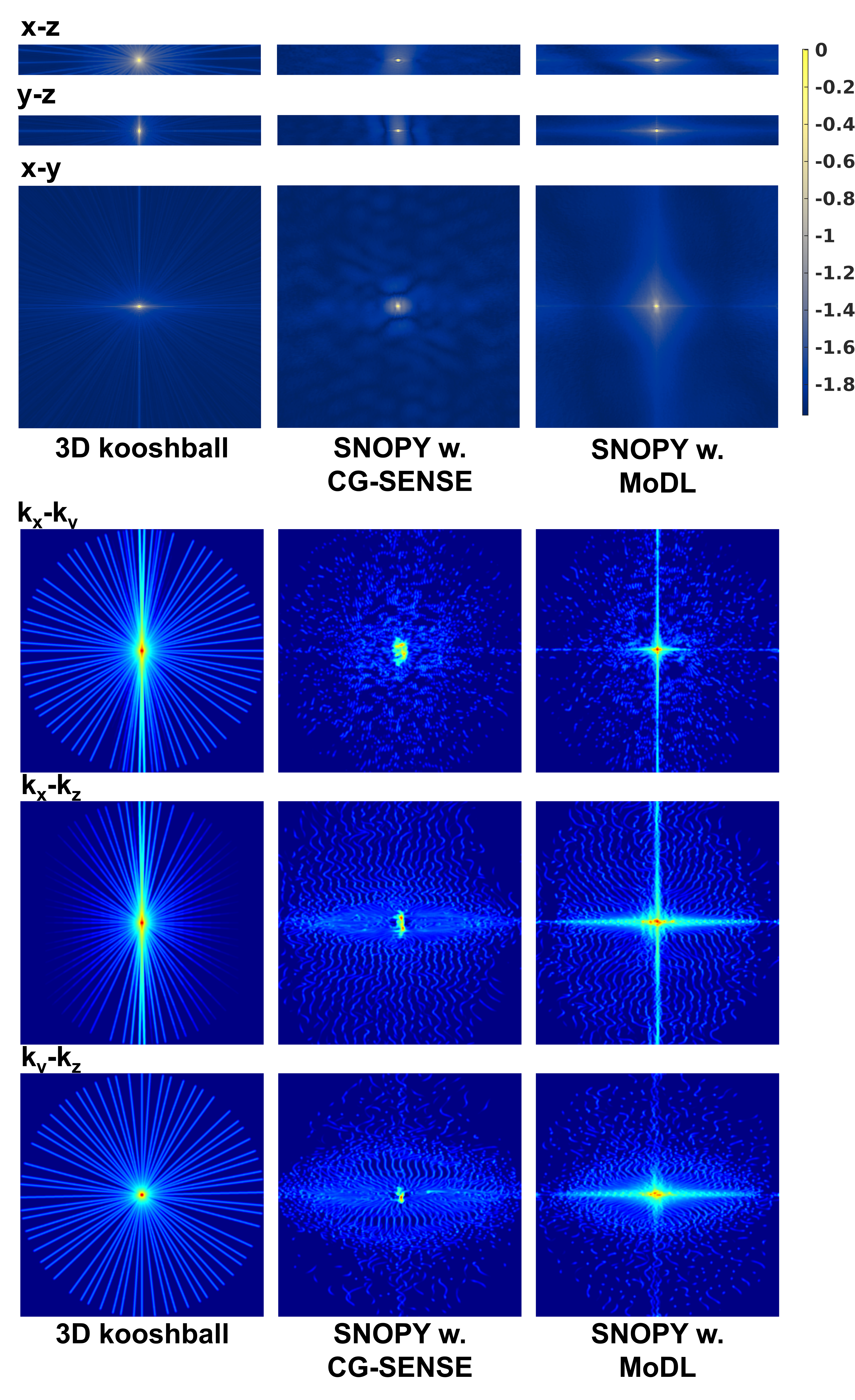}}
\caption{~Visualization of the optimized trajectory in \ref{exp:freeform}.
The upper subfigure displays PSFs (log-scaled, single-coil)
of trajectories optimized with different reconstruction methods.
The lower subfigure shows the density of sampling trajectories,
by convolving the sampling points with a Gaussian kernel.
Three rows are central profiles from three perspectives.
\label{fig:psf}}
\end{figure}

\begin{figure*}[htb]
\centerline{\includegraphics[width=0.99\textwidth]{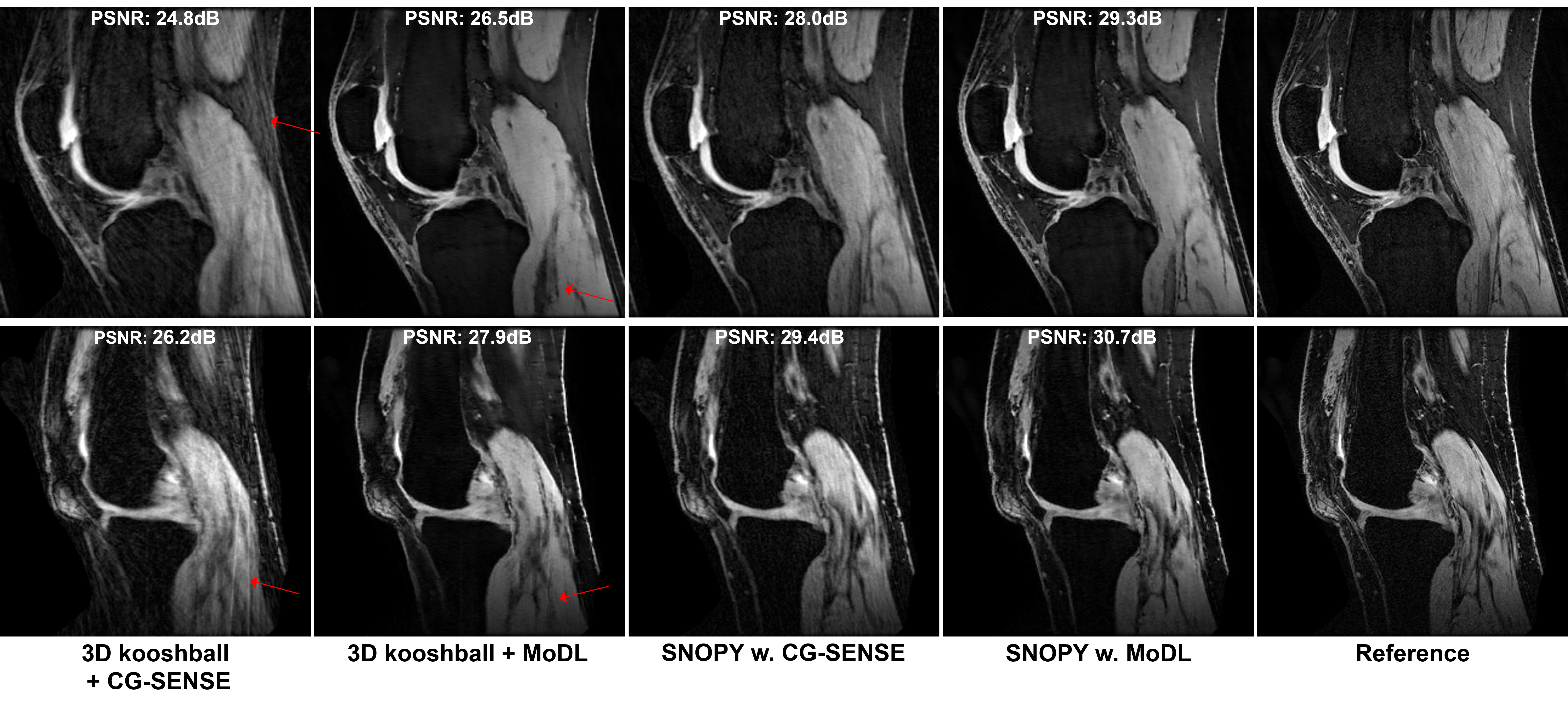}}
\caption{~Examples of the reconstructed images for two knee slices in experiment \ref{exp:freeform}.
\label{fig:recon}}
\end{figure*}

\subsubsection{Memory saving techniques}

Due to the large dimension, 
the memory cost for naive 3D trajectory optimization
would be prohibitively intensive.
We developed several techniques to
reduce memory use and accelerate training.

As discussed above,
the efficient Jacobian approximation 
uses only 10\% of the memory used in 
the standard auto-differentiation approach \cite{wang:21:eao}.
%It is also possible to
We also used in-place operations
in certain reconstruction steps,
such as the conjugate gradient (CG) method,
because with careful design
it will still permit
auto-differentiation.
(See our open-source code\footnote{\url{https://github.com/guanhuaw/Bjork}} for details.)
The primary memory bottleneck is with the 3D NUFFT operators.
We pre-calculate the Toeplitz embedding kernel to 
save memory and accelerate computation \cite{fessler:05:tbi,muckley:20:tah}.
In the training phase,
we use a NUFFT with lower accuracy,
for instance, with a smaller oversampling ratio for gridding
\cite{wang:21:eao}.
\tref{tab:memory} shows the incrementally improved efficiency
achieved with these techniques.
Without the proposed techniques,
optimizing 3D trajectories
would require hundreds of gigabytes of memory,
which would be impractical.
SNOPY enables solving this otherwise prohibitively large problem
on a single graphic card (GPU).

\section{Methods}
\label{sec:exp}
\subsection{Datasets}

We used two publicly available datasets;
both of them contain 3D multi-coil raw \kspace data. 
SKM-TEA \cite{desai2022skm}
is a 3D quantitative double-echo steady-state (qDESS\cite{welsch:2009:qdess}) knee dataset.
It was acquired by 3T GE MR750 scanners
and 15/16-channel receiver coils.
SKM-TEA includes 155 subjects.
We used 132 for training, 10 for validation,
and 13 for the test.
Calgary brain dataset \cite{souza2018open}
is a 3D brain T1w MP-RAGE \cite{brant-zawadzki:1992:MPRAGE} \kspace dataset.
It includes 67 available subjects,
acquired by an MR750 scanner
and 12-channel head coils.
We used 50 for training, 6 for validation, and 7 for testing.
All receiver coil sensitivity maps were calculated by
ESPIRiT \cite{espirit}.

\begin{table*}[bt]%
\caption{The quantitative reconstruction quality (PSNR) of the test set.
\label{tab:recon}}
\begin{tabular*}{\textwidth}{@{\extracolsep\fill}lccc@{\extracolsep\fill}}
\toprule
 & \textbf{CG-SENSE}  & \textbf{PLS}  & {\textbf{MoDL}} \\
\midrule
3D kooshball & 28.15 dB  & 28.16 dB  & 30.07 dB    \\
SNOPY & 32.47 dB  &  32.53 dB  & 33.68 dB    \\
\bottomrule
\end{tabular*}

\end{table*}

\subsection{Simulation experiments}
 
We experimented with multiple scenarios to show the broad
applicability of the proposed method. All the experiments used a server node equipped with an Nvidia Tesla A40 GPU for training.

\begin{figure*}
\centerline{\includegraphics[width=0.9\textwidth]{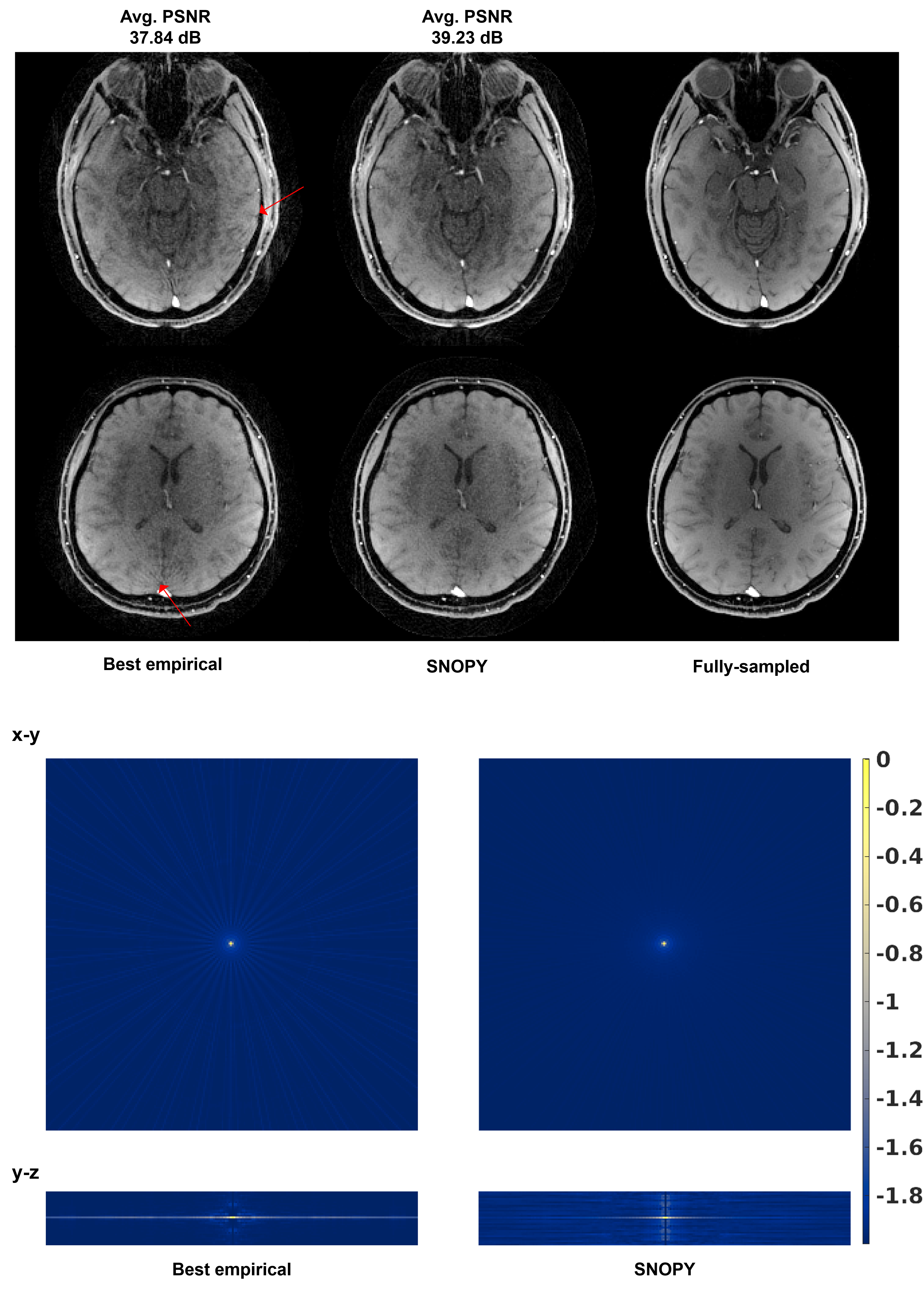}}
\caption{~Prospective results of \ref{exp:sos}, optimizing the rotation angles of the stack-of-stars (6$\times$ acceleration).
`Best empirical' uses the design from previous study \cite{zhou:2017:GoldenratioRotatedStackofstars}.
The upper subfigure shows two slices from prospective in-vivo experiments.
The reconstruction algorithm was PLS.
\textbf{Avg. PSNR} is the average PSNR of the 4 subjects
compared to the fully sampled reference.
The lower subfigure shows the log-scaled PSF (single-coil) of two trajectories. 
\label{fig:sos}}
\end{figure*}

\begin{figure*}
\centerline{\includegraphics[width=\textwidth]{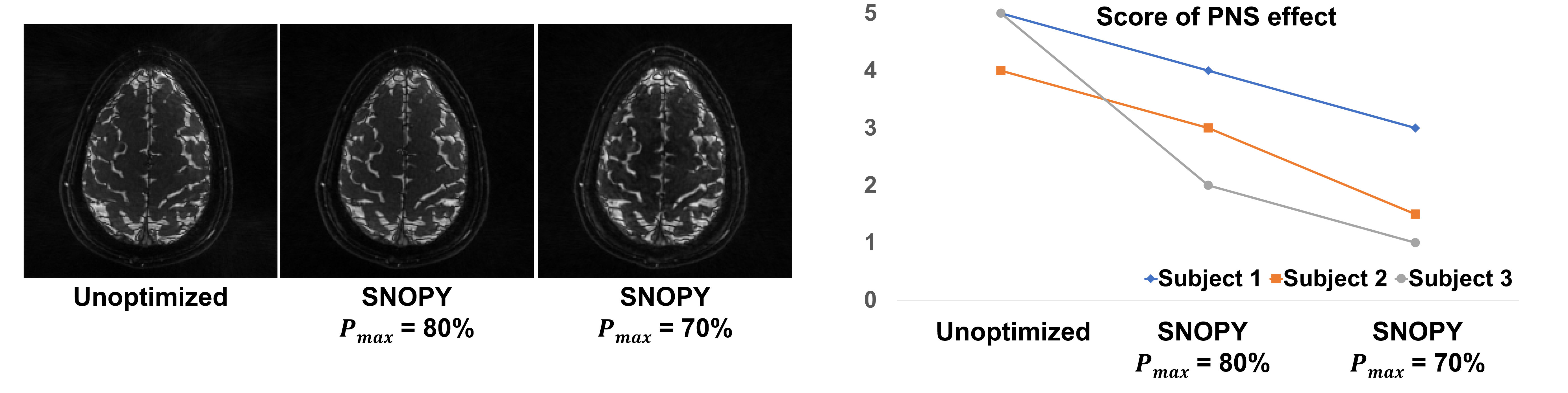}}
\caption{~Prospective results of \ref{exp:pns}. We showed three different trajectories: the unoptimized REPI, SNOPY-optimized with the PNS threshold of 80\%, and SNOPY-optimized with the PNS threshold of 70\%. The left subfigure shows one slice of reconstructed images. The reconstruction used PLS and 120 shots (volume TR = 2s). The right subfigure shows the subjective score of the PNS effect.\label{fig:ossi}}
\end{figure*}

\subsubsection{Optimizing 3D gradient waveform}
\label{exp:freeform}
We optimized the sampling trajectory with a 3D radial
(``kooshball'') initialization
\cite{barger:2002:TimeresolvedContrastenhancedImaging,herrmann:2016:TimeEfficient3D}.
As described in \ref{subsec:param},
we directly optimized the readout waveform of each shot.
The trajectory was parameterized by B-spline kernels
to reduce the number of degrees of freedom
and enable multi-scale optimization.
The initial 3D radial trajectory 
had a 5.1 2ms readout (raster time = 4 $\mu$s)
and 1024 spokes/shots (8$\times$ acceleration),
using the rotation angle described in \citestd{chaithya:2022:OptimizingFull3D}.
The training used the SKM-TEA dataset.
The FOV was 15.8$\times$15.8$\times$5.1cm
with 1mm$^3$ resolution.
The receiver bandwidth was $\pm$125kHz.
The training loss was
$$
\Loss = \Lrecon + 0.1 \Lg + 0.1 \Ls +  \Lpns.
$$
The gradient strength (\gmax),
slew rate (\smax),
and PNS threshold (\pmax)
were 50 mT/m, 150 T/m/s, 80\%,
respectively.
The learning rate
$\etao$ % macro...
decayed from 1e-4 to 0 linearly.
For the multi-level optimization,
we used 3 levels (with B-spline kernel widths = 32, 16, and 8 time samples),
and each level used 200 epochs.
The total training time was $\sim$180 hrs.
We also optimized the trajectory
for several image reconstruction algorithms.
We used a regularizer weight of 1e-3
and 30 CG iterations for CG-SENSE and PLS. 
For learning-based reconstruction,
we used the MoDL \cite{modl} approach
that alternates between a neural network-based denoiser
and data consistency updates.
We used a 3D version of the denoising network \cite{DIDN},
20 CG iterations for the data consistency update,
and 6 outer iterations.
Similar to previous investigations
\cite{aggarwal:2020:JointOptimizationSampling, wang:22:bjork-tmi},
SNOPY jointly optimized
the neural network's parameters
and the sampling trajectories
using \eqref{e:sgld}.

\subsubsection{Optimizing rotation angles of stack-of-stars trajectory}
\label{exp:sos}

This experiment optimized the rotation angles
of stack-of-stars,
which is a widely used volumetric imaging sequence.
The training used Calgary brain dataset.
We used PLS as the reconstruction method for simplicity,
with $\lambda=10^{-3}$ and 30 iterations.
We used 200 epochs and a learning rate
linearly decaying from 1e-4 to 0.
The FOV is 25.6$\times$21.8$\times$3.2 cm
and the resolution is 1mm$^3$.
We used 40 spokes per $kz$ location (6$\times$ acceleration),
and 1280 spokes in total.
The readout length is 3.5 ms.
The receiver bandwidth is $\pm$125kHz.
The trajectory was a stack of
32 stars,
so we optimized
1280
rotation angles \cc.
 
Since optimizing rotation angles
does not
impact the gradient strength, slew rate, PNS, and image contrast, 
we used only
the reconstruction loss
$\Loss = \Lrecon.$
We regard the method (RSOS-GR) proposed in previous work
\cite{zhou:2017:GoldenratioRotatedStackofstars} as the
best currently available scheme.
We applied 200 epochs with a linearly decaying learning rate 
from 1e-3 to 0.
The training cost  $\sim$20 hrs.

\subsubsection{PNS suppression of 
3D rotational EPI trajectory
for functional imaging}
\label{exp:pns}

The third application
optimizes the rotation EPI
(REPI) trajectory \cite{rettenmeier:2022:REPI},
which provides an efficient sampling strategy for fMRI.
For higher resolution (i.e., $\leq$1mm),
we found that subjects may experience
strong PNS effects introduced by REPI.
This experiment aimed to reduce the PNS effect of REPI
while preserving the original image contrast.
We optimized one shot/readout waveform of REPI
with a B-spline kernel with a width of 16 to parameterize the trajectory,
and rotated the optimized readout shot
using the angle scheme
similar to \cite{rettenmeier:2022:REPI}.

We designed the REPI readout
for an oscillating stead steady imaging (OSSI) sequence,
a novel fMRI signal model that can
improve the SNR \cite{guo:2020:OSSI,guo:2020:HOSSI}.
The FOV is 20$\times$20$\times$1.2 cm,
with 1 mm$^3$ isotropic resolution,
TR = 16 ms, and TE = 7.4 ms.
The readout length is 10.6 ms.
The receiver bandwidth is $\pm$250kHz.

To accelerate training,
the loss term here
excluded
the reconstruction loss $\Lrecon$:
$$
\Loss = 0.01 \Loss_{g} + 0.01 \Loss_{s} + \Loss_{pns} + \Loss_{c}.
$$
The training used 40,000 steps,
with a learning rate decaying linearly from 1e-4 to 0.
The training cost $\sim$1 hrs.

\subsection{In-vivo experiments}

We implemented the optimized trajectory prospectively
on a GE UHP 3.0T scanner
equipped with a Nova Medical 32-channel head coil.
Participants gave informed consent under local IRB approval. 
Because the cache in the MR system cannot load  
hundreds of distinct gradient waveforms,
the experiment \ref{exp:freeform}
was not implemented prospectively.
Readers may refer to the corresponding 2D prospective studies
\cite{wang:22:bjork-tmi}
for image quality improvement
and correction of eddy current effects.
For experiment \ref{exp:sos},
we programmed the sampling trajectory with
a 3D T1w fat-saturated GRE sequence \cite{nielsen:2018:TOPPEFrameworkRapid},
with TR/TE = 14/3.2ms and FA = 20\textdegree.
The experiment included 4 healthy subjects.
For experiment \ref{exp:pns},
to rate the PNS effect,
we asked 3 participants to score the nerve stimulation 
with a 5-point Likert scale
from `mild tingling' to `strong muscular twitch.'

\subsection{Reproducible research}

The code for 2D trajectory optimization is publicly available%
\footnote{\url{https://github.com/guanhuaw/Bjork}}.
As an accompanying project,
MIRTorch\footnote{\url{https://github.com/guanhuaw/MIRTorch}}
facilitates the differentiable programming
for MRI sampling and reconstruction.
When this paper is accepted, we will also provide the 3D version as a toolbox
on open-source platforms.
For the prospective in-vivo experiments,
we will provide open-source and vendor-agnostic 
sequences based on TOPPE \cite{nielsen:2018:TOPPEFrameworkRapid}.

\section{Results}
\label{sec:res}
For the spline-based freeform optimization experiment delineated in \ref{exp:freeform},
\fref{fig:bjork} shows an example of the optimized trajectory
with zoomed-in regions and plots of a single shot.
Similar to the 2D case \cite{wang:22:bjork-tmi}
and SPARKLING \cite{chaithya:2022:OptimizingFull3D, sparklingmrm},
the multi-level B-spline 
optimization
leads to a swirling trajectory
that can cover more \kspace in the fixed readout time,
to reduce large gaps between sampling locations
and thus help reduce aliasing artifacts.
\fref{fig:psf} displays point spread functions (PSF)
of trajectories optimized jointly with different
reconstruction algorithms.
To visualize the sampling density
in different regions of \kspace,
we convolved the trajectory with a Gaussian kernel,
and \fref{fig:psf}
shows the density 
of central profiles from different views.
Compared with 3D kooshball,
the SNOPY optimization led to fewer radial patterns in the PSF,
corresponding to fewer streak artifacts in \fref{fig:recon}.
Different reconstruction algorithms
generated distinct optimized PSFs
and densities, 
which agrees with previous studies \cite{wang:21:eao, zibetti2020fast, gozcu:2019:RSP}.
\tref{tab:recon} lists the quantitative reconstruction quality
of different trajectories.
The image quality metric is the average peak signal-to-noise ratio (PSNR) of the test set.
SNOPY led to
$\sim$4 dB higher PSNR than
the kooshball initialization.
\fref{fig:recon} includes examples of
reconstructed images.
Compared to kooshball, SNOPY's reconstructed images
have fewer artifacts and blurring.
Though MoDL reconstruction (and its variants) is one of the best reconstruction algorithms
based on the open fastMRI reconstruction challenge \cite{muckley2021results}, 
many important structures are misplaced with the kooshball reconstruction.
Using the SNOPY-optimized trajectory,
even a simple model-based reconstruction (CG-SENSE)
can reconstruct these structures.

For experiment \ref{exp:sos},
\fref{fig:sos} shows the PSF of the optimized angle 
and RSOS-GR angle scheme
\cite{zhou:2017:GoldenratioRotatedStackofstars}.
For the in-plane ($x$-$y$) PSF,
the SNOPY rotation shows noticeably reduced streak-like patterns.
In the $y$-$z$ direction, SNOPY optimization leads to 
a narrower central lobe
and suppressed aliasing.
The prospective in-vivo experiments also support this theoretical finding.
In \fref{fig:sos},
the example slices
(reconstructed by PLS) from prospective studies
show that SNOPY
reduces streaking artifacts
and blurring.
The average PSNR of SNOPY and RSOS-GR for the 4 participants 
were 39.23 dB and 37.84 dB, respectively.

\begin{figure*}
\centerline{\includegraphics[width=0.9\textwidth]{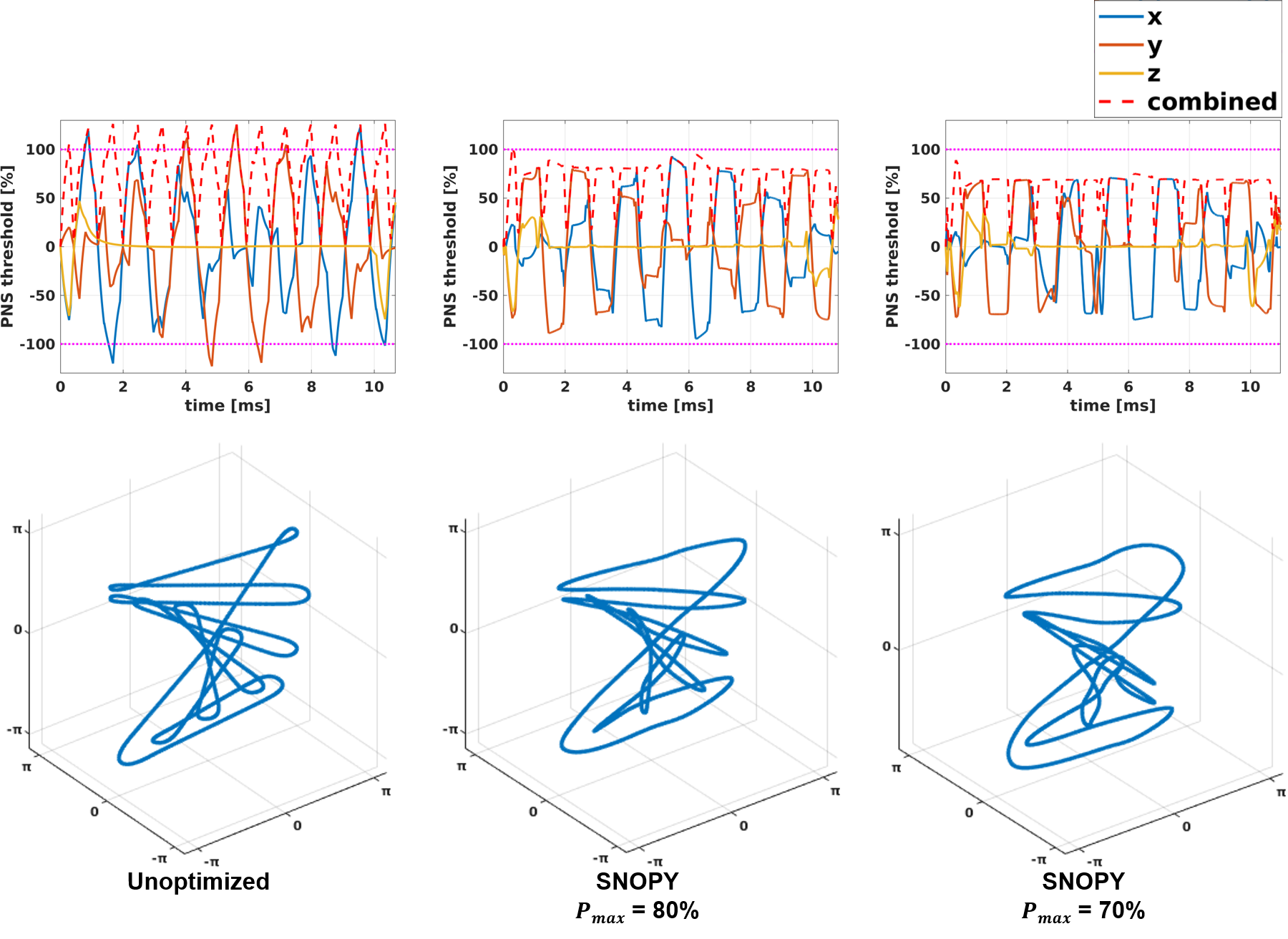}}
\caption{~The first row plots the PNS effect
calculated by the convolution model (\ref{eqn:pns}) of the experiment \ref{exp:pns}.
The second row shows one readout trajectory
before/after the SNOPY optimization.\label{fig:pns}}
\end{figure*}

In experiment \ref{exp:pns},
we tested three settings:
unoptimized REPI,
optimized with PNS threshold (\pmax in \eqref{eqn:pns}) = 80\%,
and
optimized with \pmax= 70\%.
\fref{fig:pns} shows one shot before/after the optimization,
and a plot of simulated PNS effects.
For the subjective rating 
of PNS,
the first participant reported 
5,2,1;
the second participant reported
4,3,2;
the third participant reported 5, 4, 3.
The SNOPY optimization
effectively reduced the subjective PNS effect
of the given REPI readout
in both simulation and in-vivo experiments.
Intuitively, SNOPY smooths the trajectory
to avoid a constantly high slew rate,
preventing the high PNS effect.
\fref{fig:ossi} shows one slice of
reconstructed images by the CS-SENSE algorithm.
Though SNOPY suppressed the PNS effect,
the image contrast was well preserved
by the image contrast regularizer \eqref{eqn:contrast}.

\section{Discussion}
\label{sec:discussion}
SNOPY presents a novel yet intuitive approach to optimizing
non-Cartesian sampling trajectories.
Via differentiable programming,
SNOPY enables applying gradient-based
and data-driven methods to trajectory design.
Various applications and in-vivo experiments
showed the applicability and robustness of SNOPY.

Experiments \ref{exp:freeform} and \ref{exp:sos} used training data to 
improve image quality
by trajectory optimization.
SNOPY can tailor the sampling trajectory to
specific training datasets and reconstruction algorithms
by formulating the reconstruction image quality as a training loss.
An accompanying question is
whether the learned sampling trajectories
could overfit the training dataset.
In experiment \ref{exp:sos},
the training set used an MP-RAGE sequence,
while the prospective sequence was an RF-spoiled GRE.
In a 2D experiment \cite{wang:22:bjork-tmi},
we found that trajectories learned with
one anatomy (brain), contrast (T1w),
and vendor (Siemens) still improved the image quality of
other anatomies (like the knee),
contrasts (T2w), and vendors (GE).
These empirical studies indicate that
trajectory optimization is robust to a 
moderate distribution shift between training and inference.
An intuitive explanation is that 
SNOPY can improve the PSF by reducing the aliasing,
and such improvement is universally beneficial.
In subsequent investigations
with more diverse datasets,
we plan to study the robustness of SNOPY in more settings.
For instance,
one may optimize the trajectories with healthy controls
and prospectively test the trajectories with pathological participants,
to examine the image quality of pathologies.
Testing SNOPY with different FOVs,
resolutions, and field strengths
will also be desirable.

An MRI system suffers from imperfections,
such as field inhomogeneity \cite{sutton:2003:FastIterativeImagea},
eddy currents \cite{ahn:1991:AnalysisEddycurrentInduced},
and gradient non-linearity \cite{hidalgo-tobon:2010:TheoryGradientCoil}.
Many correction approaches exist,
such as B0-informed reconstruction \cite{fessler:05:tbi}
and trajectory mapping \cite{duyn:1998:SimpleCorrectionMethod,robison:2019:CorrectionB0Eddy}.
SNOPY-optimized trajectories are compatible with these existing methods. 
For example, we implemented eddy-current correction
for a 2D freeform optimized trajectory in \citestd{wang:22:bjork-tmi}.
It is also possible to consider these perfections 
in the forward learning/optimization phase,
so the optimized trajectory
has innate robustness to imperfections.
For instance,
the forward system model \A in \eqref{eqn:image} could
include off-resonance maps.
This prospective
learning approach
will require prior knowledge of the
distribution of system imperfections,
which is usually scanner-specific and hard to simulate.
In future studies, we plan to investigate approaches
to simulate such effects prospectively.

SNOPY uses a relatively simplified model of PNS.
More precise models, such as \citestd{davids2019prediction},
may lead to improved PNS suppression results.

The training uses several loss terms,
including image quality, PNS suppression,
hardware limits, and image contrast.
By combining these terms,
the optimization can lead to trajectories that
boast multiple desired characteristics.
The weights of different loss terms were determined empirically.
One may control the optimization results by altering the coefficients.
For example, with a larger coefficient of the hardware constraint loss, the trajectory will better conform to \smax and \gmax.
Bayesian experiment design is also applicable to
finding the optimal loss weights.
Additionally,
the training losses (constraints)
may contradict each other,
and the optimization may get stuck in a local minimizer.
We considered several empirical solutions
to this problem.
Similar to SPARKLING \cite{sparklingmrm},
one may relax the constraint on maximum gradient strength
by using a higher receiver bandwidth.
Using SGLD can also help escape the local minima
because of its injected randomness.
One may also use a larger B-spline kernel width
to optimize the gradient waveform
in the early stages
of a coarse-to-fine search.

Trajectory optimization is a non-convex problem.
SNOPY uses several methods,
including effective Jacobian approximation,
parameterization, multi-level optimization, and SGLD,
to alleviate the non-convexity and lead to better optimization results.
Such methods were found to be effective in this and previous studies
\cite{wang:22:bjork-tmi,wang:21:eao}.
Initialization is also important
for non-convex problems.
SNOPY can take advantage of 
existing knowledge of MR sampling
as a benign optimization initialization.
For example,
our experiments used the well-received 
golden-angle stack-of-stars
and rotational EPI as optimization bases.
The SNOPY algorithm can continue to improve 
these skillfully designed trajectories 
to combine the best of both stochastic optimization
and researchers' insights.

SNOPY can be extended to many applications,
including dynamic and quantitative imaging.
These new applications may require 
task-specific optimization objectives in addition to
the ones described in \ref{subsec:obj}.
In particular,
if the reconstruction method
is not readily differentiable,
such as the  MR fingerprinting reconstruction
based on dictionary matching \cite{MRF},
one needs to design a surrogate objective
for image quality.

\section*{Acknowledgments}
The authors thank Dr. Melissa Haskell, Dr. Shouchang Guo, Dinank Gupta, and Yuran Zhu for the helpful discussion.

% \subsection*{Author contributions}

% \red{

% }

% \subsection*{Financial disclosure}

% None reported.

% \subsection*{Conflict of interest}

% The authors declare no potential conflict of interests.

\bibliography{ref}%
% \vfill\pagebreak
% \section*{Author Biography}

% \begin{biography}{\includegraphics[width=66pt,height=86pt,draft]{empty}}{\textbf{Author Name.} This is sample author biography text this is sample author biography text this is sample author biography text this is sample author biography text this is sample author biography text this is sample author biography text this is sample author biography text this is sample author biography text this is sample author biography text this is sample author biography text this is sample author biography text this is sample author biography text this is sample author biography text this is sample author biography text this is sample author biography text this is sample author biography text this is sample author biography text this is sample author biography text this is sample author biography text this is sample author biography text this is sample author biography text.}
% \end{biography}

% \section*{Supporting information}
% \red{
% The following supporting information is available as part of the online article:
% }

% \appendix

% \nocite{*}% Show all bib entries - both cited and uncited; comment this line to view only cited bib entries;

\end{document}